# HEP Community Summer Study 2013
# Computing Frontier Group I3: Networking

Gregory Bell (LBNL) and Michael Ernst (BNL)

## *1. Executive Summary*

Research in High Energy Physics (HEP) depends on the availability of reliable, high-bandwidth, feature-rich computer networks for interconnecting instruments and computing centers globally[i]. In fact, given the distributed data and computing models of major HEP experiments, it can be argued that research networks are becoming extensions of HEP discovery instruments.[ii]

Most HEP-related data is transported by National Research and Education Networks (NRENs), supplemented by infrastructures dedicated to specific projects. NRENs differ from commercial network providers, because they are optimized for transporting massive data flows generated by large-scale scientific collaborations. In addition, many NRENs offer advanced capabilities that commercial providers do not have an incentive to deploy.

For decades, network traffic generated by HEP has been a primary driver of NREN growth; partnerships between HEP and NREN staff have broken new ground in the field of networking; and HEP requirements have challenged NRENs and motivated their research activities. In the next ten years and beyond, the productivity of HEP collaborations will continue to depend on active partnerships between HEP and networking organizations, plus an ecosystem of innovative global NRENs.

HEP collaborations are now accustomed to viewing network transport as a reliable and predictable resource – so much so that data models for ATLAS and CMS have evolved rapidly[iii] in response to NREN capabilities – but this state of affairs is not inevitable. Other data-intensive communities have begun to generate large traffic flows and, following the example of HEP, to incorporate high-performance networks into science workflows. As a result of this broad trend toward data intensity across many disciplines, NRENs around the world will be challenged to meet the requirements of large-scale research, and must be adequately resourced in order to continue serving the critical role they have played in the past.

In support of HEP's objectives through 2020, basic and applied networking research is necessary in a range of subjects[iv]. This research should be conducted collaboratively - by a combination of network researchers, HEP community members, and NREN staff[v]. Some of the following research questions are relatively new, while others have been explored productively by members of the network research and HEP communities[1] in the past, but all will be relevant to HEP over the next ten years:

- What future architectures will maximize utilization and minimize cost in core and campus networks?
- How can emerging paradigms such as Software Defined Networking or Named Data Networking be harnessed most effectively to improve HEP science outcomes?
- Can networks evolve into adaptive, self-organizing systems – programmable at every layer – that quickly respond to requests of HEP science applications?
- If well-tuned host systems (or ensembles of them) have the ability to saturate a single backbone channel, what techniques and architectures can NRENs adopt to maximize data mobility?
- How will the emerging "complexity challenge" arising from closer integration between networks and applications be managed, especially in the multi-domain, multi-national context?
- How can diverse networks cooperate – automatically and securely – to offer science-optimized capabilities on a worldwide basis?
- Can discovery or automation techniques reduce the need for fragile, manual configuration?
- How will networks respond to the operational challenge of deploying and managing dozens of wavelengths across large geographies under relatively flat funding prospects?
- Will post-TCP protocols become useful outside of highly-controlled, "walled garden" demonstrations?
- Would computer modeling of applications, networks, and data flows be useful in answering any of these questions?
- Will power consumption become a limiting economic or operational factor in this time period?

Recent investments in network research have been too small, and continued underfunding will compromise the ability of HEP collaborations to maximize scientific productivity. Increased research funding, while necessary, is not sufficient; there also needs to be increased attention to the process of translating the results of network research into real-world architectures which NRENs can deploy and manage. Incentives and funding for such 'translational' activities are urgently needed. Because network research has now begun to intersect with research in services and applications, cross-disciplinary funding opportunities should also be available.

---

[1] Notably among them Caltech's Harvey Newman, whose impact on research networking has been longstanding and singular, plus Newman's many collaborators at Caltech, CERN, and global NRENs, national laboratories, and universities.

A number of cultural and operational practices need to be overcome in order for NRENs (and global cyber infrastructures more generally) to fully succeed: expectations for network performance must be raised significantly, so that collaborations do not continue to design workflows around a *historical* impression of what is possible; the gap between peak and average transfer rates must be closed; and campuses must deploy secure science data enclaves – or  Science DMZs[vi] – engineered for the needs of HEP and other data-intensive disciplines.  Fortunately, each of these trends is currently underway, but momentum must be accelerated.

Ten years from now, the key applications on which HEP depends will not be fully successful, efficient, or cost-effective if they are run on the Internet as it exists today. During the next decade, research networks need to evolve into *programmable instruments* – flexible resources which can be customized for particular needs, but which exist within a common, integrated, ubiquitous framework that is reliable, robust and trusted for its privacy and integrity. These are major challenges, but they are tractable if funding agencies invest in innovative research, and maintain support for the exponential growth of NREN traffic.

## *2. Fundamental forces and trends*

### *2.1 Exponential growth in network traffic*
A fundamental trend is the continuing exponential growth in HEP-related network traffic, which doubles approximately every 18 months. Although equipment upgrade cycles will inevitably produce periods of slower growth followed by step-function increases, the historical growth rates will likely continue in aggregate, because underlying Moore-law drivers (for example: detector resolution, storage capacity, computational power) continue to operate. Although national research networks have invested in 'dark fiber' infrastructures and can accommodate growth in terrestrial traffic more cost-effectively than before, trans-oceanic growth will be more challenging to accommodate, because undersea capacity is more limited and more expensive.

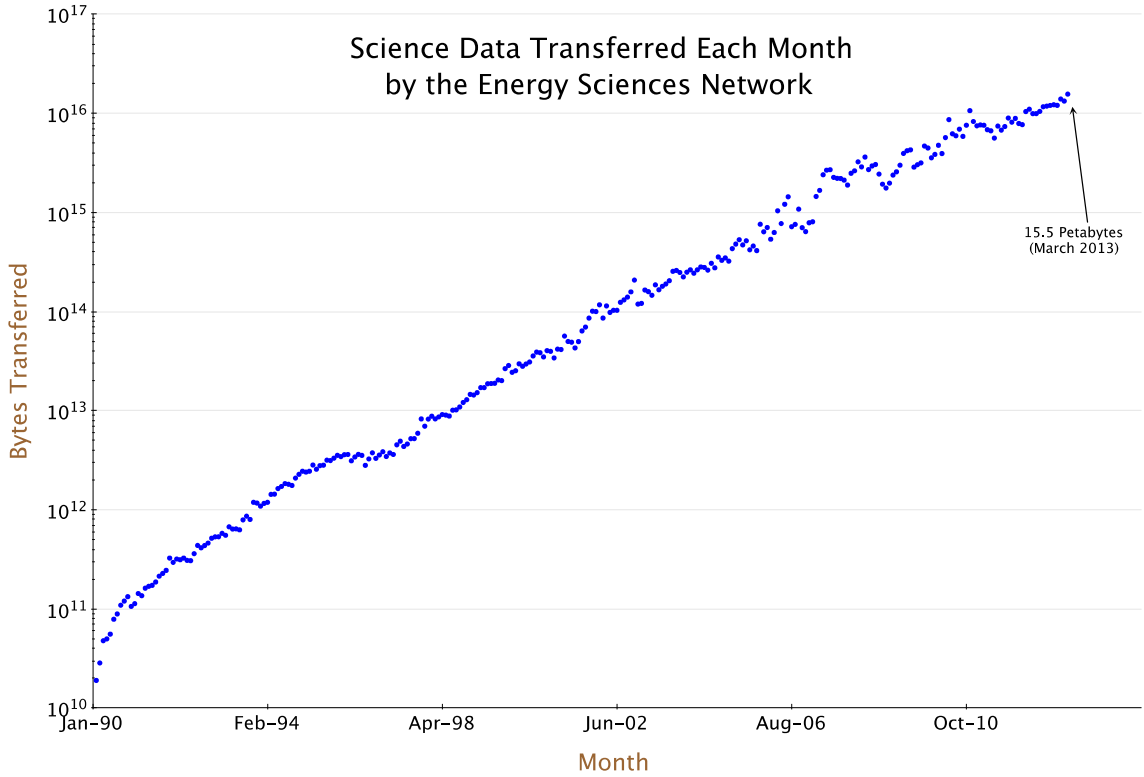

Figure 1: Network traffic transferred by DOE's Energy Sciences Network, in bytes per month, since 1990. Traffic doubles every 18 months.

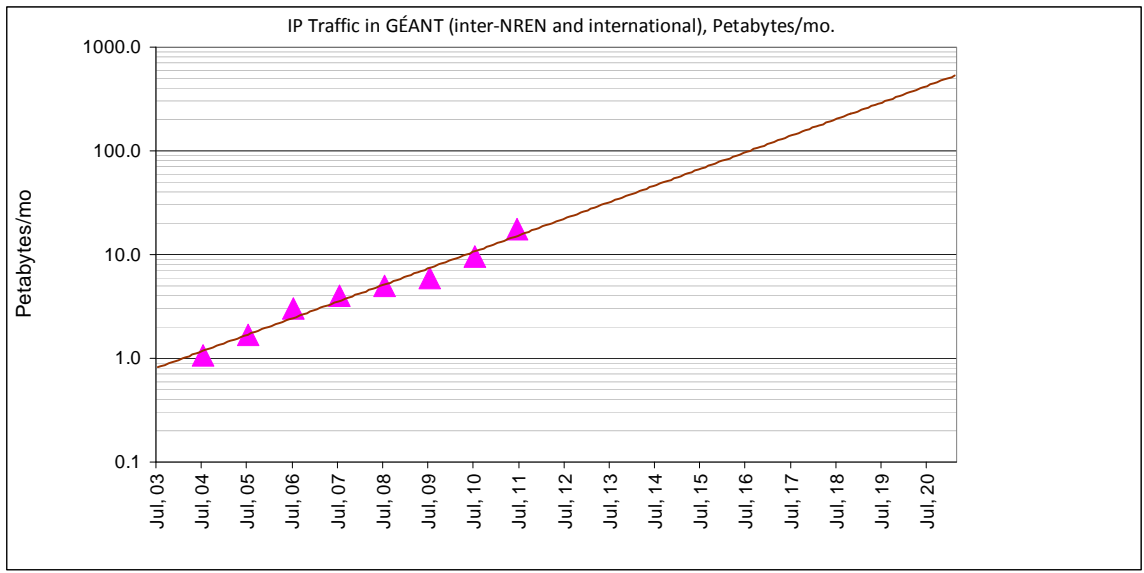

Figure 2: Historic and projected traffic volume between NRENs across the Pan-European backbone GEANT.

## 2.2 Increasing complexity

Closer integration of applications and research networks has the potential to improve science outcomes for HEP. This integration is already beginning to occur, and at the same time networks are evolving into programmable systems which can be virtualized, reserved, queried, and managed by applications and middleware. The consequence of these two trends will be a significant increase in the complexity of global networks, and this complexity will be compounded by the rapid growth described above. Managing and minimizing complexity from a global systems perspective will be an important challenge requiring significant research and development activity.

## 2.3 Pressures on sustainability

The number of range of network-connected devices is growing rapidly worldwide. In addition to millions of servers, billions of Internet users, and billions of mobile phone subscribers, we can anticipate an era in which trillions of networked sensors join the global Internet. Unless business models for network operators evolve and adapt, this development may lead to rising costs and stagnating revenues. On the current path of development, limits related to physical resources, rare materials, available spectrum, and energy will prove to be challenging.

All of these trends suggest that HEP collaborations will need to maintain even closer partnerships with global research networks to assure that scientific goals are advanced. In addition, funding agencies need to support continued innovation in networking science – especially given the emergence of new paradigms for network architecture, and the profound challenge of coordinating architectural change across many administrative domains spanning dozens of nations.

# 3. Anticipated evolution

## 3.1. Architectural evolution

There continues to be ongoing debate about the best approach for updating the global Internet architecture. Numerous proposals, both evolutionary and revolutionary, have been articulated. Some of these could be deployed incrementally, while others require a "clean slate" approach. In addition to the evolution towards a *programmable Internet* discussed above, at least one other emerging paradigm holds promise for advancing the objectives of HEP. In this "named data" or content-centric approach, applications specify the unique name rather than the logical location of a data object when fetching it from the network, and the network takes responsibility for content location and distribution.

Regardless of the extent to which SDN, NDN, or other new networking paradigms achieve broad deployment within NRENs, it is virtually certain that research networks will become more tightly integrated with each other and with science applications. As a result, they will begin to resemble a coherent global system, rather than a set of interoperating domains. One notable example of a 'virtual network' architecture already spanning multiple domains and serving the needs of LHC experiments is the highly-successful [LHCONE](#) project; LHCONE is a domain-specific network overlay serving dozens of LHC grid computing sites, and improving data mobility for many of them.

Throughout the coming decade, it can be expected that reliance on cloud computing services (access to CPU, storage, and application resources over the network) will increase.  This trend will in turn drive NRENs to peer aggressively with commercial providers, to engineer for greater uptime, and to extend science-focused capabilities - such as programmability - into cloud data centers. The extent to which this trend disrupts distributed grid computing models will depend on the economic and technical viability of large-scale cloud-sourcing for HEP applications. Because cost and service models are changing rapidly, the outcome of this process is far from clear.

## 3.2. Technological evolution

In the popular press, optical fiber has long been described as having "unlimited capacity." While it is certain that research in optical physics and materials science will produce further gains, data transmission along a given communications channel is constrained by an information-theoretical (ie, Shannon) limit.  Optical vendors have made progress in recent years developing technologies to increase transmission rates over a given band, but commercial systems will eventually approach theoretical efficiency rates, and further increases will require compromises in terms of spectral capacity or transmission distance – compromises that increase transmission costs.

This means that – in the absence of disruptive optical technology or nationwide deployment of next-generation fiber – NRENs will need to think about scaling 'horizontally' in the coming decade, deploying many high-speed optical waves in parallel, and potentially acquiring multiple fiber pairs across their national footprints. This development will have important implications on cost and manageability, and operationally focused research will be necessary to identify cost-effective and stable parallel architectures. It should be noted that other technological trends (including silicon photonics) may drive down the cost – but not the operational challenge – of horizontal scaling in the future.

For completeness, and since mobility is expected to become increasingly important for HEP computing, it should be noted that legacy radio access technologies are also approaching Shannon limits, although new (eg multi-user) paradigms promise greater spectral efficiency.

## 4. Barriers and challenges

### 4.1 Complacency (the innovator's dilemma)

In the words of the NSF workshop on Fundamental Research in Networking in 2003, "the success of the Internet carries the risk of complacency about the need for true innovation and outside-the-box thinking". In fact, substantial research is needed to investigate new Internet models and architectures, especially in the face of exponentially-increasing science traffic, constrained budgets, and the rise of network programmability. Complacency is also a problem from the perspective of user expectations: there is a significant gap between peak and average network performance. Bridging that gap will require concerted outreach and engagement on the part of the global NREN community. In the past year, there has been significant progress in deploying science-optimized campus architectures around the world, and this positive trend must be sustained through adequate funding and outreach activity.

### 4.2 Budget constraints and resource competition

Historically, HEP programs have profited from an environment of abundant network bandwidth, which other scientific communities have not been able to fully exploit. This abundance may diminish in the next few years, as more science communities make the transition to network intensity and compete for the bandwidth, capabilities, and engineering attention of NRENs. Data volume is only one dimension of the issue: global NRENs are also challenged by budget pressures, changing business models, diversity of content and services, as well as rising expectations placed on the network for reliability and security.

### 4.3. Network research is underfunded

There must be support and growth for the eroding base of basic research in areas relevant to the science of information networking. Recent investments have been too small, and the innovation dividend from prior research will not go on forever. The translation of research results into operational practices is equally critical, but poorly funded, and researchers have few incentives to build effective feedback loops between their work and those of network operators. Finally, research in network communications has begun to intersect with research in services and applications; this is a welcome trend, and more cross-disciplinary funding opportunities should be made available.

## 5. Research and innovation agenda

Future networks will integrate communication, computing, and storage resources in order to support a wide range of discovery techniques and environments. These will include domain-specific science gateways and portals, cloud-based workflows, high-performance and high-throughput computing models, and new data service capabilities.

Research and innovation in many domains are necessary to support this evolution; see the *Executive Summary* for a partial list of relevant topics. Areas of inquiry include network virtualization, programmability, application integration, content distribution, and management of complexity.

Modeling and simulation will continue to be important techniques in the toolkit of network researchers. In addition, funding for large-scale, multi-domain, multi-layer test beds – integrating network infrastructure with realistic ensembles of well-tuned systems – will be critical to assure that research insights are translated into operational reality. Finally, close cooperation with vendors will be necessary; network architectures developed for HEP should not depart substantially from commercially-available solutions, in order to exploit mass-market pricing.

[Software Defined Networking](#) and [Named Data Networking](#) hold promise as techniques for increasing the utility and reducing the cost of global networks. Other paradigms may prove equally relevant. It will be important for researcher and NREN communities to remain open to new approaches (for instance, applications of neural networking research insights), even as significant attention is justifiably focused for the moment on OpenFlow. The ultimate criterion for the value of networking research – from the perspective of HEP, at least - should be improved science outcomes.

As networks scale horizontally and integrate more tightly with science applications, control and management will become more costly. Increased automation may help reduce the burden of such complexity. The ultimate goal will be for networks to incorporate intelligence, autonomously discovering resources and adjacent systems and adjusting in real time to changes in load and topology, without human intervention. Large scale distributed control systems that are self-adaptive and self-organizing will help prevent management and operating costs from scaling exponentially with traffic load.

## *6. Summary and Outlook*

For decades, network traffic generated by HEP has been a primary driver of NREN growth; partnerships between HEP and NREN staff have broken new ground in the field of networking; and HEP requirements have challenged NRENs and motivated their research activities. In the next ten years and beyond, the productivity of HEP collaborations will continue to depend on active partnerships between HEP and networking organizations, plus an ecosystem of innovative global NRENs.

Other data-intensive communities have begun to generate large traffic flows and, following the example of HEP, to incorporate high-performance networks into science workflows. As a result, NRENs around the world will be challenged to meet the requirements of large-scale research, and must be adequately resourced and motivated.

In order for NRENs to continue meeting the needs of HEP science, numerous research questions (see *Executive Summary* for a partial list) need to be pursued. In addition, funding needs to be made available to facilitate the transition of network research into capabilities and services that are maintainable in daily operations by NRENs and their partners.

In addition, cultural change needs to occur in order for global research networks to fulfill their maximum potential. Expectations for network performance must be raised significantly, so that collaborations do not design workflows around a historical impression of what is possible. The gap between peak and average transfer rates must be closed. Finally, campuses must deploy secure, science data enclaves – or Science DMZs – engineered and optimized for the needs of data-intensive science.

---

[i] ESnet Community Network Requirements reports http://www.es.net/about/science-requirements/reports/

[ii] W. Johnston, E. Dart, M. Ernst, B. Tierney „Enabling high throughput in widely distributed data management and analysis systems: Lessons from the LHC", http://www.es.net/assets/pubs_presos/High-throughput-lessons-from-the-LHC-experience.Johnston.TNC2013.pdf, Terena Network Conference June 2013

[iii] I. Fisk "Computing Model Evolution to Validation", http://indico.cern.ch/getFile.py/access?contribId=2&sessionId=0&resId=0&materialId=slides&confId=251191, WLCG Collaboration Workshop, November 2013

[iv] I. Monga "Software-Defined Networks – Bridging the Application-Network Divide", http://indico.cern.ch/getFile.py/access?contribId=485&sessionId=0&resId=1&materialId=slides&confId=214784, Conference in Computing in High Energy and Nuclear Physics 2013

[v] H. Newman "Advanced Networking for HEP, Research and Education in the LHC era" http://indico.cern.ch/getFile.py/access?contribId=484&sessionId=0&resId=1&materialId=slides&confId=214784, Conference in Computing in High Energy and Nuclear Physics 2013

[vi] E. Dart, B. Tierney "The Science DMZ: A Network Design Pattern for Data-Intensive Science", http://www.es.net/assets/pubs_presos/sc13sciDMZ-final.pdf,